\documentclass{article} %
\usepackage{iclr2025_conference,times}

\usepackage{amsmath,amsfonts,bm}
\usepackage{xspace}

\newcommand{\ie}{i.e.,\xspace} %

\newcommand{\sk}{k}%
\newcommand{\nonce}{\eta}%
\newcommand{\PRNG}{\operatorname{PRNG}}
\newcommand{\Enc}{\operatorname{Encr}}
\newcommand{\Dec}{\operatorname{Decr}}
\newcommand{\WM}{\operatorname{WM}}
\newcommand{\GS}{\operatorname{GS}}
\newcommand{\mes}{s}
\newcommand{\ciph}{c}
\newcommand{\adv}{\mathcal{A}}
\newcommand{\negl}{\operatorname{negl}}
\newcommand{\gen}{\mathcal{G}}
\newcommand{\inv}{\mathcal{I}}
\newcommand{\samp}{\mathcal{S}}
\newcommand{\secparam}{n}
\newcommand{\game}[1]{\mathsf{PrivK}^{\mathsf{IND\$-CPA}}_{\mathcal{A},#1}(\secparam)}
\newcommand{\gameprime}[1]{\mathsf{PrivK}^{\mathsf{IND\$-CPA}}_{\mathcal{A}',#1}(\secparam)}

\newcommand{\x}{\ensuremath{x}\xspace}  %
\newcommand{\z}{\ensuremath{z}\xspace}  %
\newcommand{\invz}{\ensuremath{\hat{\z}}\xspace}  %
\newcommand{\zT}{\ensuremath{\z_T}\xspace}  %
\newcommand{\zThat}{\ensuremath{\invz_T}\xspace}  %

\newcommand{\repl}{\ensuremath{\operatorname{Repl}}\xspace}  %

\newcommand{\s}{\ensuremath{m}\xspace}  %
\newcommand{\sd}{\ensuremath{\mes}\xspace} %
\newcommand{\shat}{\ensuremath{\hat{m}}\xspace}  %

\def\eqref#1{equation~\ref{#1}}

\def\1{\bm{1}}

\DeclareMathAlphabet{\mathsfit}{\encodingdefault}{\sfdefault}{m}{sl}
\SetMathAlphabet{\mathsfit}{bold}{\encodingdefault}{\sfdefault}{bx}{n}

\usepackage{hyperref}
\usepackage[capitalise, noabbrev]{cleveref}
\usepackage{graphicx}
\usepackage{tikz}
\usepackage{booktabs}
\usetikzlibrary{shapes.geometric, arrows, positioning}
\tikzset{
    box/.style={draw, rectangle, rounded corners, text width=4cm, minimum height=5cm, align=left},
    arrow/.style={thick,->,>=stealth},
}

\usepackage{url}

\title{Towards A Correct Usage of Cryptography in Semantic Watermarks for Diffusion Models}

\author{Jonas Thietke, Andreas Müller, Denis Lukovnikov, Asja Fischer\thanks{Equal supervision}~, Erwin Quiring\footnotemark[1]~\\
Ruhr University Bochum, Germany \\
\texttt{\{firstname.lastname\}@rub.de}
}

\iclrfinalcopy %
\begin{document}

\maketitle

\begin{abstract}
Semantic watermarking methods enable the direct integration of watermarks into the generation process of latent diffusion models by only modifying the initial latent noise. One line of approaches building on Gaussian Shading relies on cryptographic primitives to steer the sampling process of the latent noise. However, we identify several issues in the usage of cryptographic techniques in Gaussian Shading, particularly in its proof of lossless performance and key management, causing ambiguity in follow-up works, too. In this work, we therefore revisit the cryptographic primitives for semantic watermarking. We introduce a novel, general proof of lossless performance based on IND\$-CPA security for semantic watermarks. We then discuss the configuration of the cryptographic primitives in semantic watermarks with respect to security, efficiency, and generation quality. 
\end{abstract}

\section{Introduction}
Inversion-based semantic watermarks are a novel class of watermarking methods for latent diffusion models (LDMs)~\citep{Wen2023TreeRing, Yang2024GaussianShading, CiYanSon24RingID, gun2024prc}. 
These watermarks change the initial latent noise to contain a watermark pattern which is recovered later by inverting the denoising process in a diffusion model. Hence, the diffusion model does not need to be changed. Semantic watermarks are thus easy to implement and empirical results suggest a high robustness to image perturbations. %
As only the initial latent is changed, the watermark %
becomes a plausible, inherent part of the generated image, for instance, through specific object details.%

The semantic watermarking methods differ in how they modify the initial latent noise~\zT and can be categorized into two types. %
\emph{Distribution-changing} methods such as Tree-Ring~\citep{Wen2023TreeRing} and RingID~\citep{CiYanSon24RingID} add fixed circular patterns into the frequency spectrum of \zT which changes the distribution of \zT. In contrast, \emph{distribution-preserving} methods such as Gaussian Shading~\citep{Yang2024GaussianShading} and PRC~\cite{gun2024prc} keep the distribution of~\zT. They critically rely on \emph{cryptographic primitives} to generate a pseudorandom sequence that steers the sampling process~of~\zT. 

However, Gaussian Shading, laying the foundation for distribution-preserving semantic watermarking, has not properly specified the cryptographic primitives. This can cause considerable ambiguity on the usage of this watermark method. %
In particular, we find that the proof of lossless performance, proving the watermark's undetectability, is not accurately specified. Moreover, it only covers the scenario where each user generates only a single image ever.
The implications for key management are not discussed either, leading to multiple possible configurations of Gaussian~Shading.

In this work, we revisit the cryptographic primitives for semantic watermarking. We focus on Gaussian Shading, but also discuss PRC.
First, we present a novel, general proof based on IND\$-CPA security \citep{indscpa} that allows demonstrating the lossless performance of a semantic watermark. It also covers the realistic watermarking scenario where a user generates multiple images.
We apply this proof on Gaussian Shading and discuss its applicability to PRC.

Second, we analyze the implications of this proof for the key management in semantic watermarking regarding security, efficiency, and generation quality and variety.  
We show that the secure configuration of Gaussian Shading does not affect the generation quality and variety, but leads to a rather inefficient scheme in terms of runtime and storage. PRC, in contrast, can also be deployed efficiently.

\section{Background} %
In this section, we shortly describe the applied cryptographic principles and the watermarking process of Gaussian Shading. Without loss of generality, we focus on the multi-bit scenario where the watermark allows detection and user identification.

\paragraph{Stream Ciphers.}
One of the key advances in Gaussian Shading is the use of a stream cipher to guide the image generation process. In general,
a stream cipher is an encryption algorithm that encrypts a message~$\mes$ by generating a keystream which is combined with the message to create a ciphertext that can be securely transmitted. 
In a nutshell, the cipher works as follows.
First, a pseudo-random generator ($\PRNG$) is used to obtain a keystream,
$K = \PRNG(\sk, \nonce)$, where $\sk$ is a secret key and $\nonce$ is a public nonce.
Afterwards, the message is encrypted using bitwise XOR: \mbox{$\ciph = K \oplus \mes$}.
Note that $\ciph$ looks like a random bit string, which is an essential property for its use in Gaussian Shading.
In order to decrypt~$\ciph$, the receiver first uses $\PRNG$ to obtain the same keystream~$K$ as used for encryption, and then obtains the original message by using bitwise XOR: $\mes = K \oplus \ciph$.
In summary, the encryption function is $\Enc(\sk, \nonce,\mes) = \PRNG(\sk, \nonce) \oplus \mes$, and the decryption function is $\Dec(\sk, \nonce,\ciph) = \PRNG(\sk, \nonce) \oplus \ciph$.
We refer to \cite{katz} for a more elaborate explanation.
A secure stream cipher should ensure that every change of even one bit in $\sk$ and $\nonce$ entirely changes the output of $\PRNG$. Otherwise, various attacks would be possible~\citep{symmetricCrypto2}. 
Gaussian Shading applies ChaCha20~\citep{bernstein2008chacha} as stream cipher, which requires a 256-bit secret key $\sk$ and a 96-bit nonce~$\nonce$.

\paragraph{Watermark Generation And Verification.}
Figure~\ref{fig:gs-overview} in Appendix~\ref{sec:appendix-gs} illustrates the watermarking process of Gaussian Shading~\citep{Yang2024GaussianShading}.
Before generating images, the provider generates a random user id \s for each user.
When the user prompts the provider to generate a new image given some textual description, the provider first samples a latent $z_T$ with a sampling strategy $\samp$ and subsequently uses it to generate a new image $x$ using the generator $\gen$ through iterative denoising starting from $z_T$.
Gaussian Shading is integrated in the first step by changing the default sampling strategy based on a standard Gaussian $\mathcal{N}(O, I)$.
Gaussian Shading instead uses the user identifier $\s$ to steer the sampling of $z_T$.
First, \s is replicated several times to increase robustness during message recovery, yielding $\sd = \repl(\s)$.
The replicated user id~\sd is then encrypted using a stream cipher: $\ciph=\Enc(\sk,\nonce,\sd)$.
This encrypted message $\ciph$ is used to steer the sampling procedure~$\samp$.
To this end, the standard normal distribution is divided into $2^\ell$ bins, each with equal probability mass.
The elements of $\ciph$ are used to select the bins of $\mathcal{N}(0, 1)$ from which each element of the random vector $z_T$ is sampled.
For $\ell=1$, we have two bins and randomly sample either a negative or a positive value $\zT[i]$ from the Gaussian distribution depending on the binary value of~$\ciph[i]$. %
In summary, we can describe the process to obtain a watermarked image~$\x$ as a concatenation of multiple functions: 
$\x = \GS(\sk,\nonce,\s) = \gen(\samp(\Enc(\sk, \, \nonce, \, \repl(\s))))$.

For watermark verification of an image~$\x'$, the model provider performs a full inversion~$\inv$ to get an estimated latent noise~$\zThat~=~\inv(\x')$. Next, the inverse sampling process~$\samp^{-1}$ is done where~$\zThat$ is quantized to obtain the encrypted message bits~$\hat{\ciph}$. 
After decrypting $\hat{\ciph}$ with $\Dec(\sk,\nonce,\hat{\ciph})$ and applying error correction with $\repl^{-1}$, we obtain the recovered user id $\shat$. Taken together, this process can be described as follows:
\mbox{$\shat = \GS^{-1}(\sk,\nonce,\x') = \repl^{-1}(\Dec(\sk, \, \nonce, \, \samp^{-1}(\inv(\x'))))$.}
The final stage is to check if $\shat$ matches with any user id~$\s$ known by the service provider.
This is done by comparing the number of matched bits between $\shat$ and every known $\s$.
A match is found if the number of matching bits exceeds a pre-defined threshold.

\section{Revisiting Cryptography for Semantic Watermarking}
We proceed with a critical assessment of the cryptographic principles used in Gaussian Shading. As described earlier, a symmetric stream cipher is used to create a pseudorandom message that controls the sampling process. This has several advantages. 
The distribution of $\ciph$ becomes uniform, so that it can be shown that the sampling process still results in a Gaussian distribution~\citep{Yang2024GaussianShading}.
Moreover, it enables proving undetectability. \citeauthor{Yang2024GaussianShading} show that the watermark created by Gaussian Shading does not introduce any pattern in the image by providing \emph{a proof of lossless performance}. It builds on a security definition from steganography and states the following: If there is no polynomial time algorithm that can tell if an image is watermarked without having the secret key, then there cannot be any patterns in the image.
This means it is not possible to detect the watermark by a simple pattern recognition algorithm.

However, we identify several issues. The definition made by Gaussian Shading is an unclear specification of \cite{stegosec}. %
Neither does it cover the actual usage of watermarking with multiple images that are generated and need to be watermarked. 
Furthermore, the key management is not specified, causing a considerable ambiguity regarding the deployment of Gaussian Shading, which has already affected further work building on Gaussian Shading (see~Section~\ref{sec:key-management}). 

In the following, we address these cryptographic shortcomings of Gaussian Shading with the aim to provide general insights on how cryptography should be used for semantic watermarking.
In particular, we first propose to use IND\$-CPA security \cite{indscpa} as a more rigorous security definition and show how to establish the undetectability of semantic watermarks within this framework. We apply this proof to Gaussian Shading and outline its applicability to PRC.
Second, we examine the implications for the key management and discuss how a semantic watermark needs to be deployed to fulfill the undetectability proof. We discuss both Gaussian Shading and PRC.

\subsection{A Novel Proof of Lossless Performance}
\label{sec:losslessproof}
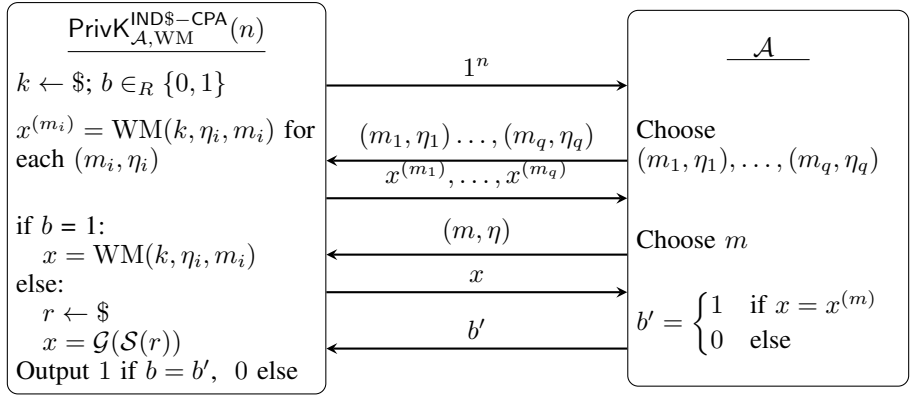
\begin{figure}
    \centering
\begin{tikzpicture}[node distance=3cm and 4cm]

    \node[box] (game) {
        \centering \underline{$\game{\WM}$}\\[0.2cm]
        \raggedright$\sk \leftarrow \$ $;
        $b \in _R \{0,1\}$\\[0.2cm]
        $x^{(\s_i)} = \WM(\sk,\nonce_i,\s_i)$
        for each $(\s_i,\nonce_i)$\\[.5cm]

        if $b$ = 1:\\
        \quad $x =  \WM(\sk,\nonce_i,\s_i)$\\
        else:\\
        \quad $r \leftarrow \$ $\\
        \quad $x = \gen(\samp(r))$\\
        Output $1$ if $b=b'$,~ $0$ else
    };

    \node[box, right=4cm of game, text width=3.5cm] (attacker) {
    		\centering\underline{ $\ \ \mathcal{A}\ \ $ \ }\\[0.7cm]
        \raggedright Choose $(\s_1,\nonce_1), \dots, (\s_q,\nonce_q)$\\[0.7cm]
        Choose $m$\\[0.5cm]
        $b' = \begin{cases} 
          1 & \text{if } x = x^{(\s)} \\ 
          0 & \text{else}
          \end{cases}$
        };

    \def\xOffset{2}

    \draw[arrow] ([yshift=1.5cm] game.east) -- ++(\xOffset,0) node[pos=1.0, above] {$1^n$} -- ([yshift=1.5cm] attacker.west);
    
       \draw[arrow] ([yshift=0.5cm] attacker.west) -- ++(-\xOffset,0) node[pos=1.0, above] {$(\s_1,\nonce_1)\dots,(\s_q,\nonce_q)$} -- ([yshift=0.5cm] game.east);

    \draw[arrow] ([yshift=0cm] game.east) -- ++(\xOffset,0) node[pos=1.0, above] {$x^{(\s_1)},\dots,x^{(\s_q)}$} -- ([yshift=0cm] attacker.west);

    \draw[arrow] ([yshift=-0.75cm] attacker.west) -- ++(-\xOffset,0) node[pos=1.0, above] {$(\s,\nonce)$} -- ([yshift=-0.75cm] game.east);

    \draw[arrow] ([yshift=-1.25cm] game.east) -- ++(\xOffset,0) node[pos=1.0, above] {$x$} -- ([yshift=-1.25cm] attacker.west);
    
    \draw[arrow] ([yshift=-2cm] attacker.west) -- ++(-\xOffset,0) node[pos=1.0, above] {$b'$} -- ([yshift=-2cm] game.east);
\end{tikzpicture}
 \caption{IND\$-CPA game for a watermarking scheme $\WM$}
    \label{fig:indscpa}
\end{figure}

In cryptography, a standard security assumption is IND-CPA (Indistinguishability under Chosen Plaintext Attack). Informally, this means that an adversary~$\adv$ cannot determine which message was encrypted, even if $\adv$ knows the two possible messages and has observed encryptions of other (not necessarily different) messages before~\citep{katz}.
IND\$-CPA is a slightly stronger assumption, stating that $\adv$ cannot distinguish the encryption of a known message from a random bit string without knowing the key. Formally, we define \textbf{the game} $\game{\WM}$, which is a polynomial time algorithm and depicted in \cref{fig:indscpa}. It uses the watermarking algorithm $\WM(\sk,\nonce,\s)$ to output a watermarked image using the secret key and a nonce which introduces randomness. 
In the beginning, it generates a random secret key $\sk$ and transmits the security parameter as $1^\secparam$ to $\adv$. 

In the \textbf{first phase} of the game, the adversary can request up to $q$ watermarked images for messages and nonces $(\s_i, \nonce_i)$ that $\adv$ provides. The inputs can be identical or different. The game responds with the corresponding watermarked images $x^{(\s_i)}$. 
In the \textbf{second phase}, the adversary provides a message $\s$ and a nonce $\nonce$. Based on the random bit $b$, \x is returned, which is either an image containing $\s$ as watermark or an image generated from an initial latent $\samp(r)$ for a distribution-preserving sampler $\samp$ and a random seed $r$. Finally, $\adv$ outputs a guess~$b'$ if \x is watermarked or not; and the game checks whether this is correct ($b = b'$).
Overall, a watermarking scheme is called IND\$-CPA-secure with a security parameter $\secparam$, if $\Pr[\game{\WM}=1]=\frac{1}{2}+ \negl(\secparam)$, where $\negl(\secparam)$ is a negligible function, \ie $2^{-\secparam}$.

An adversary~$\adv$ is called \textbf{nonce-respecting} if $\adv$ never queries the same nonce multiple times. Note that a practical adversary usually has no control over the nonce if a protocol is designed securely.
In case $\adv$ is nonce-respecting, we can show that Gaussian Shading is secure and undetectable in our new definition.
We assume that ChaCha20 is IND\$-CPA secure, as no non-generic attacks are known so far\footnote{Note that no real world symmetric cipher fulfills that definition in a strict sense. Nevertheless, the best known attacks require $\mathcal{O}(2^{\secparam/2})$ time \cite{symmetricCrypto2} which we consider infeasible for any practically relevant attack. Therefore, it can only guess a key with negligible success probability or needs to observe the encryption for one of its $q$ requested nonces.}.
Formally, we get
\begin{align}
    \Pr[\game{\GS} = 1] &= \Pr[b=1]\Pr[\adv(x^{(\s)})=1] +  \Pr[b=0]\Pr[\adv(x)=0]\\
    &= \frac{1}{2}\Pr[\adv(\GS(\sk,\nonce,\s))=1] + \frac{1}{2}\Pr[\adv(\gen(\samp(r))) = 0]\\
    \intertext{On a real random input (second term), $\adv$ cannot obtain any information and therefore just guesses with probability~$\frac{1}{2}$. On a watermarked input (first term), $\adv$ needs to recognize the output of ChaCha20 for an unknown key, which is hard by assumption. Therefore, the right hand side gets}
    &=\frac{1}{2}(\frac{1}{2}+(q+1)\negl(\secparam)) + \frac{1}{2}\cdot\frac{1}{2} = \frac{1}{2}+\negl(n) \enspace.
\end{align}

Note that \citeauthor{Yang2024GaussianShading} show this behavior for an adversary~$\adv$ that has $q=0$ queries. This adversary cannot use the same nonce multiple times as only one image is seen.
However, in practice, users and thus attackers can generate multiple images. 
Our new definition holds for this case.%

If we consider an adversary $\adv'$ that is \textbf{not nonce-respecting}, there is an obvious attack. First, $\adv'$ chooses $\s^\star$ and $\nonce^\star$ and passes this as $(\s_1,\nonce_1)$ and as $(\s,\nonce)$. $\adv'$ obtains $x^{(\s_1)}$ and $x$.
Next, $\adv'$ uses inversion\footnote{Note that $\adv'$ could even use a proxy model for inversion instead of the original model \citep{muller2024black}.} and the inverse sampler to recover the ciphertexts $\ciph_1 = \samp^{-1}(\inv(x^{(\s_1)}))$ and $\ciph = \samp^{-1}(\inv(x))$. Idealized, if they both match\footnote{Usually, they will not be exactly the same due to error in the recovery. However, they are close enough such that recovery is possible, \ie $\ciph \approx \ciph_1$.}, $\adv'$ has found that this image is watermarked and outputs $1$, otherwise $0$. We compute the probability for this attacker and find that
\begin{align}
    \Pr[\gameprime{\GS} = 1] &= \Pr[b=1]\Pr[\adv'(x^{(\s)})=1] +  \Pr[b=0]\Pr[\adv'(x)=0]\\
    &= \frac{1}{2}\Pr[\adv'(\GS(\sk,\nonce,\s))=1] + \frac{1}{2}\Pr[\adv'(\gen(\samp(r))) = 0]\\
    &= \frac{1}{2}\cdot(1) + \frac{1}{2} (1 - \negl(n)) \\
    &= 1 - \frac{1}{2}\negl(n)
\end{align}
Clearly, $\adv'$ has a non-negligible success probability---which is in fact close to $1$ even with just one watermarked image---and can therefore easily distinguish between an unwatermarked image and a watermarked one.
However, if the watermarked image is distinguishable from an unwatermarked one, given previous watermarked images, then the distribution of consecutively generated watermarked images differs from the original distribution obtained from unwatermarked $z_T$'s.

In summary, if we do not choose a new nonce for every generated image, it is easy to distinguish these images from a random one as their latents are highly similar. However, if we alter the nonce for every image, the watermark stays hidden. %

Note that PRC~\citep{chr2024prccrypto} fulfills the same security definition. It incorporates the nonce in the generation of their encrypted message $\ciph$ in the generation process in a specified way and draws a new nonce for every image.

\subsection{Key Management}
\label{sec:key-management}
Our novel proof of lossless performance extends the watermark usage to the realistic application scenario where multiple images are watermarked. The proof also shows how the encryption parameters $\sk$ and $\nonce$ need to be specified---which has not been done for Gaussian Shading in its original publication. 
In the following, we compare the recommended cryptographic configuration with other configurations of how Gaussian Shading is currently deployed.   
In addition to security considerations, we also consider the practical deployment in terms of efficiency and generation quality\,/\,variety. In the latter case, we empirically assess the impact of each configuration in Table~\ref{tab:watermark_comparison}. Our empirical setup is described in Appendix~\ref{sec:appendix-quality-variety}. In our analysis, we also shortly discuss the configuration of PRC. 

As our proof shows, the stream cipher needs to be used in a \textbf{same key, new nonce} configuration\footnote{Note that the \textbf{new key, same nonce} configuration is equivalent, as nonce and secret key are interchangeable as far as our analysis is concerned.}. This means the provider creates a fixed key~$\sk$ once. Given a fixed user id~\s and the fixed key~$\sk$, the provider has to use a new nonce~$\nonce$ to control the sampling process for every generated image.  
This configuration fulfills the security definition from Section~\ref{sec:losslessproof} and is the way how to securely use a semantic watermark such as Gaussian Shading. 
Table~\ref{tab:watermark_comparison} also shows that watermarking in this configuration preserves the quality and variety of the generated images compared to pure diffusion without any watermark. 
However, from a storage and runtime perspective, this watermarking setup is quite inefficient as it does not scale with the number of users and generated images / used nonces. 
In a normal message exchange setting, the sender could transmit the unencrypted nonce together with the encrypted message $\ciph$. 
Unfortunately, we cannot just append $\nonce$ to $\ciph$, as we need to transmit it in a robust fashion and the redundancy is applied before. Repeating the bits after the encryption would make the scheme non-random and allow for an easy distinction in IND\$-CPA security.
Hence, the provider has to store every used nonce for all the images it ever generated and, for watermark verification, has to XOR the encrypted message~$\hat{\ciph}$ with every possible keystream, which in turn requires a decryption with every stored nonce. %
PRC~\citep{gun2024prc} solves the nonce problem of Gaussian Shading by embedding its nonce into the watermarked image in a robust way.

An equivalent way in the semantic watermarking setup is the \textbf{new key, new nonce} configuration. This is in fact how Gaussian Shading is implemented in the GitHub repository~\citep{GSgithubLastCommit}. This configuration has no cryptographic benefits compared to the previous configuration, and only increases storage requirements due to the need to save every key in addition.

Finally, there is the \textbf{same key, same nonce} configuration. 
This was the configuration that the follow-up work PRC~\citep{gun2024prc} assumed for its Gaussian Shading baseline. 
This configuration is efficiently deployable, but clearly not secure. 
It is not nonce respecting, so that $\adv'$ can easily distinguish between marked and unmarked images. Moreover, this configuration of Gaussian Shading reduces image quality and variability (see Table~\ref{tab:watermark_comparison}). As the inputs to encryption and sampling are always identical for each user, the initial latent noise vectors remain similar as well.

\begin{table}[]
\small
\centering
\begin{tabular}{@{}llll@{}}
\toprule
Configuration                                           & FID $\downarrow$   & CLIP $\uparrow$            & LPIPS $\uparrow$               \\ \midrule
No Watermark                                            & $66.7459_{0.6793}$ & $0.3311_{0.0387}$          & $0.6295_{0.0812}$              \\
Same key, new nonce                                    & $66.6537_{0.7014}$ & $0.3311_{0.0400}$          & $0.6274_{0.0744}$              \\ 
Same key, same nonce                                  & $68.9729_{0.5114}$ & $0.3314_{0.0393}$          & $0.5472_{ 0.0823}$             \\
\bottomrule
\end{tabular}
\caption{Comparison of image quality and variety of different Gaussian Shading configurations. Compared to pure generation without watermark, our recommended \textit{same key, new nonce} configuration preserves quality and variety. Note that the \textit{new key, new nonce} setup would lead to equivalent results and is thus omitted. See Figure~\ref{fig:variety} in Appendix~\ref{sec:appendix-quality-variety} for additional images examples.}
\label{tab:watermark_comparison}
\end{table}

\section{Conclusion}
In this work, we revisit the cryptographic primitives of distribution-preserving semantic watermarks. We provide a novel, more general proof to show that a semantic watermark has a lossless performance. The proof also covers the realistic case where multiple images are generated. We also properly specify the encryption parameters and discuss the implications regarding security, efficiency, and image quality \& variety. 
In summary, a semantic watermark needs to be deployed such that a new nonce is used for every generated image. 

Unfortunately, this configuration makes Gaussian Shading rather inefficient to deploy. 
In contrast, PRC fulfills our novel proof too and can be efficiently deployed.
Finally, we note that it has implemented Gaussian Shading in the unfavorable configuration that affects both the security and generation quality. A direction comparison between PRC and Gaussian Shading is still open.

\bibliographystyle{iclr2025_conference}
\bibliography{ConferencesDefinition, bibliography}

\appendix

\section{Watermarking Process in Gaussian Shading}
\label{sec:appendix-gs}
\begin{figure}[h!]
    \centering
    \includegraphics[width=0.75\linewidth]{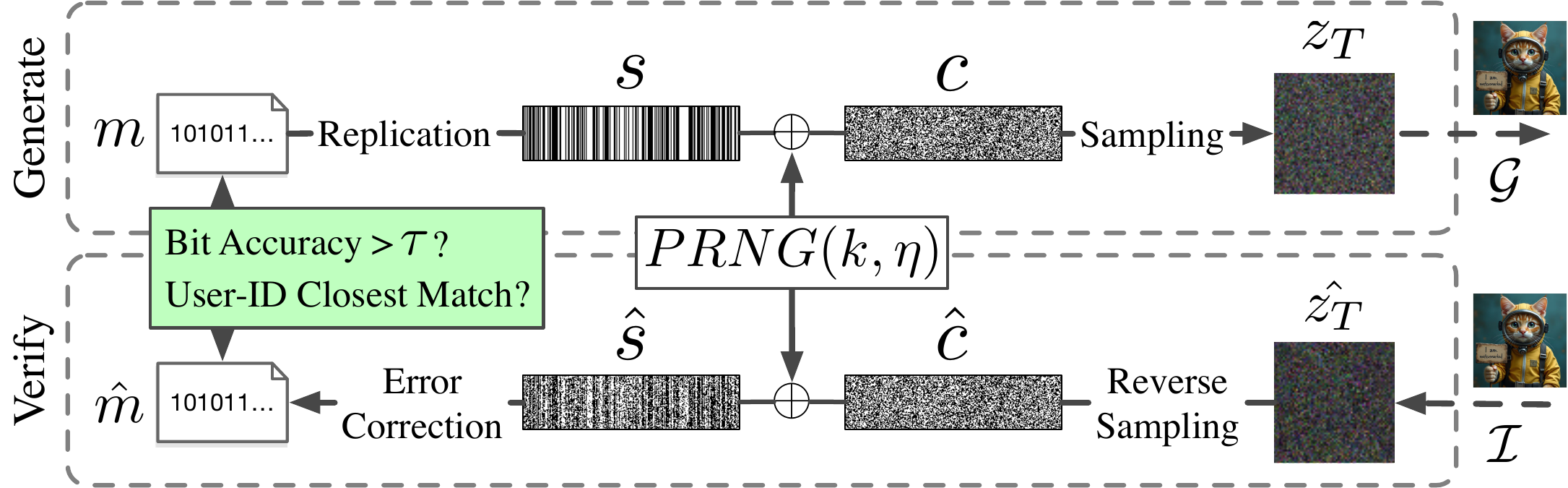}
    \caption{
    Overview over the watermark generation and verification process in Gaussian Shading
    }
    \label{fig:gs-overview}
\end{figure}

Figure~\ref{fig:gs-overview} shows the generation and verification process in Gaussian Shading.

\section{Image Quality and Variety}
\label{sec:appendix-quality-variety}

\paragraph{Setup.}
To examine the impact of each Gaussian Shading configuration on image quality and variety, we use a setup similar to~\citet{gun2024prc}. We use Stable Diffusion 2.1\footnote{\href{https://huggingface.co/stabilityai/stable-diffusion-2-1-base}{Stable Diffusion 2.1 Huggingface model card}} to generate $1.000$ images from MS-COCO~\cite{coco} validation dataset captions and calculate FID~\citep{heu2017fid} with their ground truth images five times and report the mean value and standard deviation. We further report CLIP Scores~\citep{rad2021clip} across all 5,000 generated images and their respective captions.
In order to assess the variability of images, we generate 100 images from 10 diverse prompts from the prompthero website\footnote{\href{https://prompthero.com/}{PromptHero webpage}} and calculate pairwise LPIPS~\citep{zhang2019lpips} scores, similarly to the evaluation in~\citep{gun2024prc}.
We use the DPM sampler\footnote{\href{https://huggingface.co/docs/diffusers/main/en/api/schedulers/multistep_dpm_solver}{Huggingface DPMSolverMultistepScheduler documentation}}, as well as default guidance scale (7.5) and inference steps (50).

\begin{figure}[h!]
    \centering
    \includegraphics[width=0.75\linewidth]{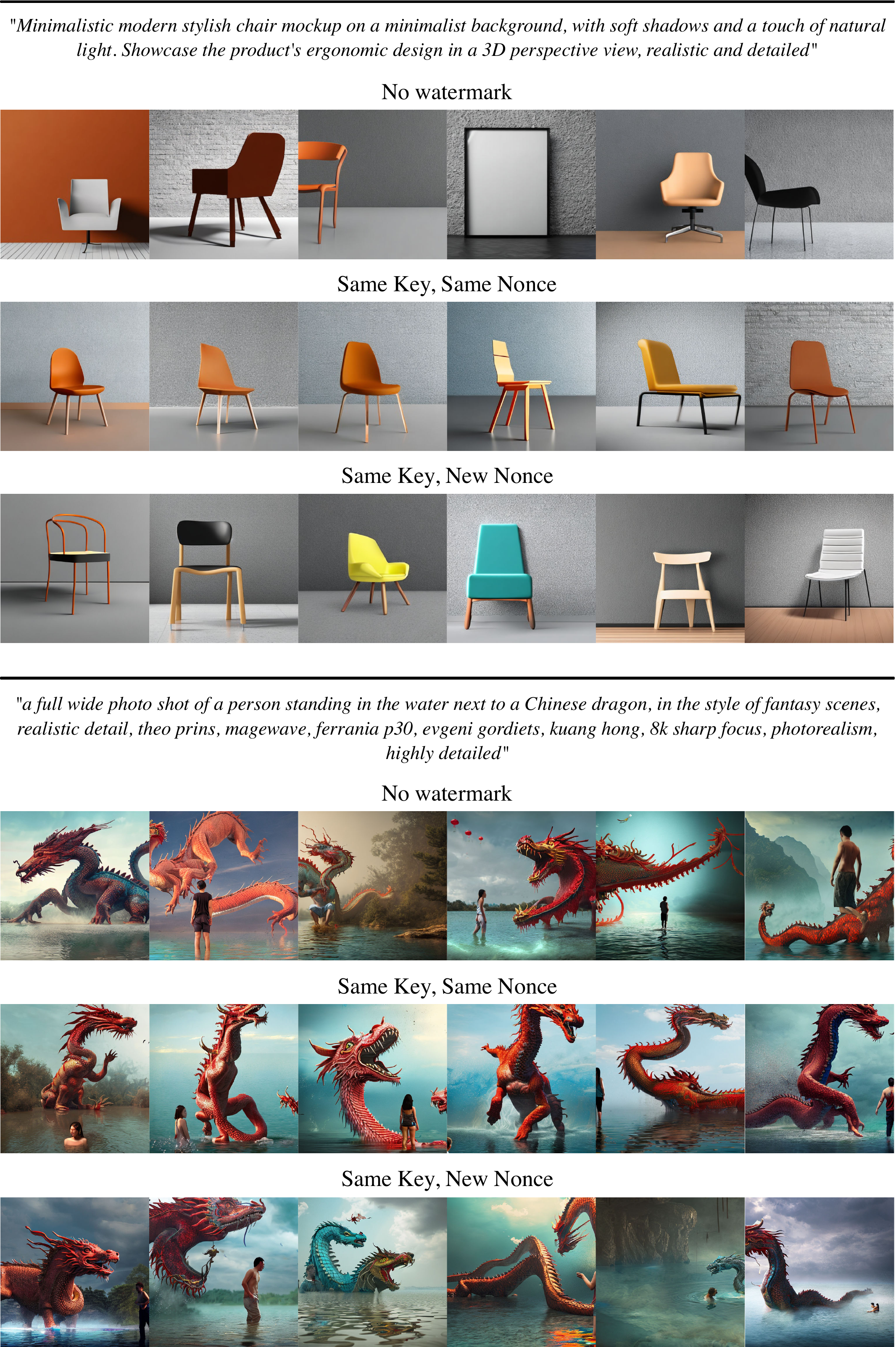}
    \caption{
    Impact of Gaussian Shading configurations on image variety.
    }
    \label{fig:variety}
\end{figure}

\paragraph{Visual Examples.}
Figure~\ref{fig:variety} shows the impact of Gaussian Shading configurations on image variety. While the same key, same nonce configuration leads to images with similar layouts, drawing new nonces for each image restores the image variety seen without watermarking.

\end{document}